\documentclass[aps,prd,onecolumn,groupedaddress,showpacs,nofootinbib,amssymb
]{revtex4}
\usepackage[dvips]{graphicx}
\usepackage{amssymb}
\usepackage{amsmath}
\usepackage{graphicx}
\usepackage{amsfonts}
\usepackage{bm}

\begin{document}

\title{Gauss-Bonnet Cosmology Unifying Late and Early-time Acceleration Eras with Intermediate Eras}

\author{V.~K.~Oikonomou$^{1,2}$}
\email{v.k.oikonomou1979@gmail.com, voiko@sch.gr}
\affiliation{ $^{1)}$Laboratory for Theoretical Cosmology, Tomsk State University of Control Systems
and Radioelectronics (TUSUR), 634050 Tomsk, Russia}
\affiliation{ $^{2)}$National Research Tomsk State University, 634050 Tomsk,
Russia}

\begin{abstract}
In this paper we demonstrate that with vacuum $F(G)$ gravity it is possible to describe the unification of late and early-time acceleration eras with the radiation and matter domination era. The Hubble rate of the unified evolution contains two mild singularities, so called Type IV singularities, and the evolution itself has some appealing features, such as the existence of a deceleration-acceleration transition at late times. We also address quantitatively a fundamental question related to modified gravity models description of cosmological evolution: Is it possible for all modified gravity descriptions of our Universe evolution, to produce a nearly scale invariant spectrum of primordial curvature perturbations? As we demonstrate, the answer for the $F(G)$ description is no, since the resulting power spectrum is not scale invariant, in contrast to the $F(R)$ description studied in the literature. Therefore, although the cosmological evolution can be realized in the context of vacuum $F(G)$ gravity, the evolution is not compatible with the observational data, in contrast to the $F(R)$ gravity description of the same cosmological evolution. 
\end{abstract}


\maketitle

\section{Introduction}

The observational data coming from Type IA supernovae in the late 90's \cite{riess} indicated that in addition to the radiation and matter domination eras, there is also a late-time acceleration era, which occurred after the matter domination era and the Universe is expanding in an accelerating way ever since. This observation utterly changed our perception for the Universe cosmological evolution, and it is a challenge for theoretical cosmologists to consistently describe the evolution of the Universe. With regards to the radiation domination era, at the beginning of this era it is believed that the Universe experienced a rapid acceleration era known nowadays as the inflationary era. During this era the Universe increased its size significantly and after the end of this era, the Universe continued its expansion in a decelerating way, with the effective equation of state of the Universe being described by a radiation perfect matter fluid. It is therefore compelling to find a model which can harbor all the evolution era of the Universe, in the same theoretical framework. Modified gravity \cite{reviews11,reviews12,reviews13,reviews14,reviews15,reviews16}, in all its aspects and variant models, can describe consistently the evolution of the Universe, so it is one of the most appealing candidates for the cosmological theory of everything, at least in the context of classical evolution. Recently we have demonstrated in Ref. \cite{nooa}, that it is possible to describe in a unified way all the cosmological eras of the Universe. The vital feature of the cosmological evolution which we realized, was the appearance of two Type IV singularities occurring at the end of the inflationary era and at the end of the matter domination era respectively. The resulting cosmological model had quite appealing features, since near the second Type IV singularity, the deceleration-acceleration transition occurred. Also, the Hubble radius decreased in a rapid way at early times, then it started to increase during the radiation and matter domination era, and finally it started to decrease at late times. The behavior of the Hubble radius is exactly what is expected from a correct description of the Universe evolution, since the primordial quantum modes exited the horizon at early times and reentered the horizon during the expansion of the Hubble radius. In addition to this, we demonstrated that the model generates a nearly scale invariant power spectrum of primordial curvature perturbations. 

Having in mind that the study performed in Ref. \cite{nooa} was realized in the context of an $F(R)$ gravity, in this paper we will realize the cosmological evolution of Ref. \cite{nooa}, but in the context of $F(G)$ gravity \cite{sergegauss1,sergegauss2,sergegauss3,sergegauss4,fg4,fg4a,fg5,fg6,fg7,fg8}, see also Refs. \cite{reviews11,reviews12,reviews13,reviews14,reviews15} for reviews. In $F(G)$  gravity, instead of the Ricci scalar $R$, used in Einstein-Hilbert and $F(R)$ gravity, the Gauss-Bonnet invariant is used, that is $G=R^2-4R_{\mu \nu}R^{\mu \nu}+R_{\mu \nu \rho \sigma}R^{\mu \nu \rho \sigma}$, where $R_{\mu \nu}$, denotes the Ricci tensor, while $R_{\mu \nu \rho \sigma}$ denotes the Riemann tensor. The basic questions we want to address with this study are the following: Firstly, what is the $F(G)$ gravity which realizes the cosmological evolution of Ref. \cite{nooa} and secondly, is this cosmology compatible with the latest Planck \cite{planck1,planck2} observations? Particularly, we will investigate if the resulting power spectrum is scale or nearly scale invariant, as was in the $F(R)$ gravity case? The answer to the second question is not obvious and with our study we will quantitatively answer this question. As we demonstrate, the resulting spectrum is not scale invariant, and hence this shows that even it is possible to realize the same cosmological evolution by using different theoretical frameworks, the observational implications of each model is not necessarily compatible with observational data. For similar studies using the framework of modified gravity in order to unify all acceleration eras of the Universe, see \cite{sergnoj,Nojiri:2006gh,capp}.
  
The singularity structure of the model we shall study contains no crushing singularities, since it contains two Type IV singularities. The latter ones are the mildest singularities, firstly classified in Ref. \cite{Nojiri:2005sx}, and recently intensively studied in \cite{Barrow:2015ora,noo1,noo3,noo4,noo5,noome}. The first novel study on mild singularities was performed in Ref. \cite{barrownew}, and later on these were developed in \cite{Barrow:2004xh,Barrow:2004hk} and see also \cite{barrowsing1,barrowsing2,barrowsing3,szy}. The interesting feature of the Type IV singularities is that it is possible for the Universe to smoothly pass through these, without any catastrophic implications on the physical quantities that can be defined on the three dimensional spacelike hypersurface defined at the time instance that the singularity occurs. However, the mild singularities affect significantly the dynamics of the cosmological evolution, with the most interesting case being the possibility that mild singularities can generate graceful exit from inflation \cite{noo5}.

We need to stress that the cosmological model which we shall study does not describe the pre-inflationary era, where a cosmic singularity at $t=0$ might occur. In fact, in the literature there exist various cosmological scenarios for this pre-inflationary era, in which case a crushing singularity might or might not occur, see for example \cite{piao1,piao2,piao3,piao4}. However, in our case the focus will be on later times and particularly for times near the beginning of inflation and after.

This paper is organized as follows: In section II we present all the essential features of the unified cosmological model, and particularly how it succeeds to describe the early-time acceleration era, the late-time acceleration era, the radiation domination era and the matter domination era. We also describe in brief the origin of primordial curvature perturbations and the behavior of the Hubble horizon. In section III we investigate how the unification evolution can be described by a vacuum $F(G)$ gravity and in section IV, using well known techniques \cite{inflation1,inflation2,inflation3,inflation4,inflationtsont,inflation5,inflation6,inflation7,inflation8,inflation9}, we calculate in detail the resulting power spectrum of the primordial curvature perturbations and the corresponding spectral index. In addition, we investigate how the primordial perturbations evolve after the horizon crossing. The conclusions follow in the end of the paper.

\section{A Brief Description of the Unified Cosmological Evolution}

In this section we describe in some detail the cosmological model which was developed in Ref. \cite{nooa}, in order to understand the qualitative features of the cosmological evolution described by it. As we now demonstrate, the model describes in a unified way the inflationary era, in terms of an initial quasi-de Sitter evolution, the radiation domination and matter domination era, and the late-time acceleration era. As we shall see, certain fine-tuning of the free parameters is needed so that a late-time acceleration era is achieved. The Hubble rate of the unified cosmological evolution is appears below \cite{nooa}, 
\begin{equation}\label{newmodel}
H(t)=e^{-(t-t_s)^{\gamma }} \left(\frac{H_0}{2}-H_i (t-t_i)\right)+f_0 |t-t_0|^{\delta } |t-t_s|^{\gamma }+\frac{1}{\sqrt{3}}
\frac{e^{\tanh (t-t_m)\ln \sqrt{\frac{4}{3}}}}{t+\frac{1}{H_0}}\, ,
\end{equation}  
where the parameter $t_m$ is chosen to be of the order $t_m\simeq 10^{12}$sec and it characterizes the transition from the radiation domination to the matter domination era. Also the parameter $t_0$ is assumed to characterize the late-time era and also $t_s$ characterizes the early-time era. Particularly, at the time instances $t=t_s$ and $t=t_0$, finite time singularities occur, which depending on the values of the parameters $\gamma$ and $\delta$, can be one of the following types,
\begin{itemize}\label{lista}
\item If $\gamma,\delta<-1$, then the cosmological evolution develops two Big Rip singularities.
\item If $-1<\gamma,\delta<0$, then the cosmological evolution develops two Type III singularities.
\item When $0<\gamma,\delta<1$, then the cosmological evolution develops two Type II singularities.
\item When $\gamma,\delta>1$, then the cosmological evolution develops two Type IV singularities.
\end{itemize} 
It is conceivable that certain combinations can occur, if the parameters $\gamma$ and $\delta$ are chosen appropriately, for example we can have a Big Rip singularity at late times and a Type IV singularity at early times, if $\gamma>1$ and $\delta<-1$, but we shall assume that both $\gamma$ and $\delta$ are chosen as $\gamma ,\delta>1$. So this means that two Type IV singularities occur for the cosmological model (\ref{newmodel}). Particularly, the time instance $t=t_s$ is assumed to occur at early times and at the end of the inflationary era, while $t=t_0$ is assumed to occur at late times, near the deceleration-acceleration transition. In order to avoid certain instabilities which might be generated for certain values of $\gamma$ and $\delta$ \cite{noo5,noome}, we further assume that $\gamma,\delta>2$. 

Also we assume that the parameters have the following values,
\begin{equation}\label{parameterschoicegeneral}
\gamma=2.1,\,\,\,\delta=2.5,\,\,\,t_0\simeq 10^{17}\mathrm{sec},\,\,\,t_s\simeq 10^{-15}\mathrm{sec},\,\,\,H_0\simeq 6.293\times 10^{13}\mathrm{sec}^{-1},\,\,\,H_i\simeq 6\times 10^{26}\mathrm{sec}^{-1},\,\,\,f_0=10^{-96}\mathrm{sec}^{-\gamma-\delta-1}\, ,
\end{equation}
and as we show the parameter $f_0$ crucially determines the late-time behavior of the model. For the parameters chosen as in (\ref{parameterschoicegeneral}), the cosmological evolution of the model (\ref{newmodel}) at early times is described by the first term of (\ref{newmodel}). When we refer to the early times, it is meant that early times correspond to the inflationary era and right after that, in which case, the function $\tanh( t-t_m)$, for $t_m=10^{12}$sec, and for $t\ll t_m$, is approximately equal to,
\begin{equation}\label{tanhearly}
\tanh(t-t_m)\simeq \tanh (-tm)\simeq -1\, .
\end{equation}
Consequently, the last term of the Hubble rate (\ref{newmodel}) is approximated at early times as follows,
\begin{equation}\label{approlastterm}
\frac{1}{\sqrt{3}}\frac{e^{\tanh (t-t_m)\ln \sqrt{\frac{4}{3}}}}{t+\frac{1}{H_0}}\simeq \frac{1}{\sqrt{3}}\frac{e^{-\ln \sqrt{\frac{4}{3}}}}{t+\frac{1}{H_0}}\simeq \frac{H_0}{2}\, ,
\end{equation}
owing to the fact that during the slow-roll inflationary phase, we have $t\ll \frac{1}{H_0}$. Consequently, the Hubble rate of the cosmological evolution at early times is approximated by a quasi de Sitter evolution of the form,
\begin{equation}\label{approxearlynew}
H(t)\simeq H_0-H_i (t-t_i)\, .
\end{equation}
The effective equation of state (EoS) of the cosmological evolution, which for modified gravity models is equal to \cite{reviews11},
\begin{equation}\label{eos}
w_{\mathrm{eff}}=-1-\frac{2\dot{H}}{3H^2}\, ,
\end{equation}
takes the following form at early times,
\begin{equation}\label{earlytimeseos}
w_{\mathrm{eff}}=-1-\frac{2 H_i}{3 \left(\frac{H_0}{2}+H_i (t-t_i)\right)^2}\simeq -1-\frac{8 H_i}{3 H_0^2}\, ,
\end{equation}
where we took into account that $t\ll 1$. Since the term $\frac{8 H_i}{3 H_0^2}\ll 1$, the EoS corresponds to a nearly de Sitter evolution at early times, since $w_{\mathrm{eff}}\simeq -1$. 

After the inflationary era, which ends at approximately $t\simeq 10^{-15}$sec, the Universe is described by the radiation domination era which will last until $t=\simeq 10^{12}$sec, as we now demonstrate. For the values of the cosmic time chosen in the interval $10^{-12}\mathrm{sec}<t<10^{10}$sec, the function  $\tanh (t-t_m)$ is approximately equal to $\tanh (t-t_m)\simeq -1$, and owing to the fact that $t\gg \frac{1}{H_0}$, the only dominant term is the Hubble rate (\ref{newmodel}) is the last term, which is approximated by,
\begin{equation}\label{approxlastnew}
\frac{1}{\sqrt{3}}\frac{e^{\tanh (t-t_m)\ln \sqrt{\frac{4}{3}}}}{t+\frac{1}{H_0}}\simeq \frac{1}{\sqrt{3}}\frac{e^{-\ln \sqrt{\frac{4}{3}}}}{t+\frac{1}{H_0}}\simeq \frac{1}{\sqrt{3}}\frac{e^{-\ln \sqrt{\frac{4}{3}}}}{t}\simeq \frac{1}{2t}\, .
\end{equation}
Since in this case, $H(t)\simeq \frac{1}{2t}$, clearly this corresponds to a radiation domination era, with the EoS being equal to $w_{\mathrm{eff}}=\frac{1}{3}$. The behavior around the cosmic time $t=10^{12}$ is a bit peculiar, since it describes an intermediate era between the radiation domination and matter domination era. In this case, for $10^{11.5}\mathrm{sec}<t<10^{12.5}$sec, the function $\tanh (t-t_m)$ is approximately equal to zero, that is, $\tanh (t-t_m)\simeq 0$, and therefore, the Hubble rate (\ref{newmodel}) becomes,
\begin{equation}\label{approxlastnew}
H(t)\simeq \frac{1}{\sqrt{3}}\frac{e^{\tanh (t-t_m)\ln \sqrt{\frac{4}{3}}}}{t+\frac{1}{H_0}}\simeq \frac{1}{\sqrt{3}t}\, ,
\end{equation}
which corresponds to an EoS approximately equal to, $w_{\mathrm{eff}}\simeq -1+\frac{2\sqrt{3}}{3}\simeq 0.15$. Clearly, the value of the EoS $0<w_{\mathrm{eff}}=0.15<\frac{1}{3}$ describes an intermediate era between the matter and radiation era, and it is conceivable that the EoS continuously deforms from $w=1/3$ to $w=0$. In some sense, this could describe a new matter form, which is collisional \cite{collisional}, that is, weakly self-interacting. After the radiation to matter domination transition, and in the cosmic time interval $10^{13}\mathrm{sec}<t<10^{17}$sec, the function $\tanh(t-t_m)$ is approximated by $\tanh (t-t_m)\simeq 1$, so the Hubble rate is approximated by the following expression,
\begin{equation}\label{approxlastnew}
H(t)\simeq \frac{1}{\sqrt{3}}\frac{e^{\tanh (t-t_m)\ln \sqrt{\frac{4}{3}}}}{t+\frac{1}{H_0}}\simeq \frac{1}{\sqrt{3}}\frac{\ln \sqrt{\frac{4}{3}}}{t+\frac{1}{H_0}}\simeq \frac{2}{3t}\, ,
\end{equation}
which corresponds to a matter domination era, with an EoS $w_{\mathrm{eff}}\simeq 0$. Finally, after the time instance $t_0=10^{17}$, the second term of the Hubble rate (\ref{newmodel}) starts to dominate, so this term generates the late-time acceleration era, with an EoS parameter,
\begin{equation}\label{esosasexeto}
w_{\mathrm{eff}}=-1-\frac{2 t^{-1-\gamma -\delta } (\gamma +\delta )}{3 f_0}\, ,
\end{equation}
which for $t\gg t_0$ describes a nearly de Sitter acceleration era. 

\begin{figure}[h] \centering
\includegraphics[width=16pc]{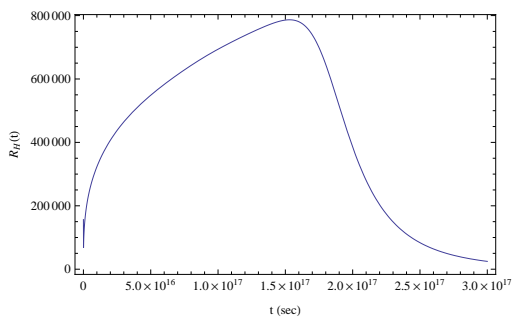}
\caption{The Hubble radius $R_H(t)$ as a function of the cosmic time, for $\gamma=2.1$ $\delta=2.5$, $t_0\simeq 10^{17}\mathrm{sec}$, $t_s\simeq 10^{-15}\mathrm{sec}$, $H_0\simeq 6.293\times 10^{13}\mathrm{sec}^{-1}$, $H_i\simeq 6\times 10^{26}\mathrm{sec}^{-1}$, $f_0=10^{-96}\mathrm{sec}^{-\gamma-\delta-1}$, $t_m=10^{12}$sec.}
\label{hubradiusnew}
\end{figure}

\begin{figure}[h] \centering
\includegraphics[width=14pc]{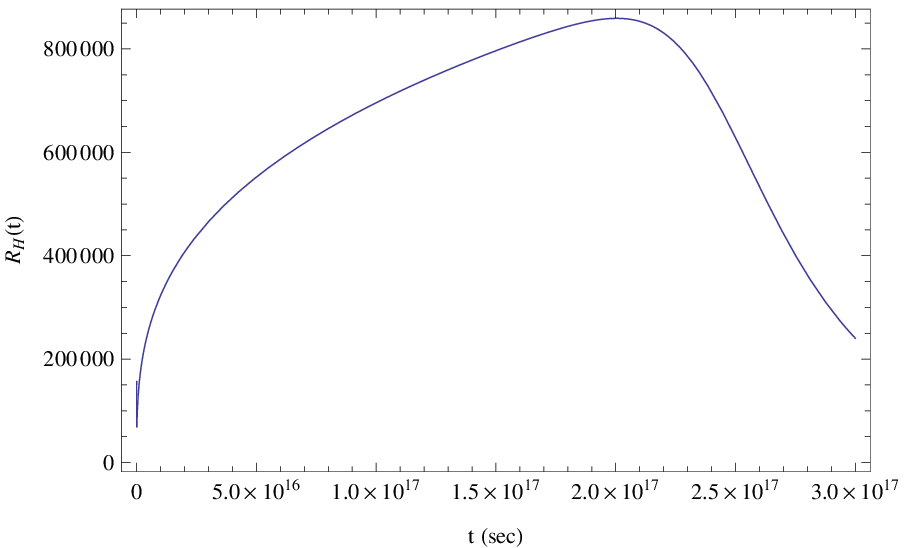}
\includegraphics[width=14pc]{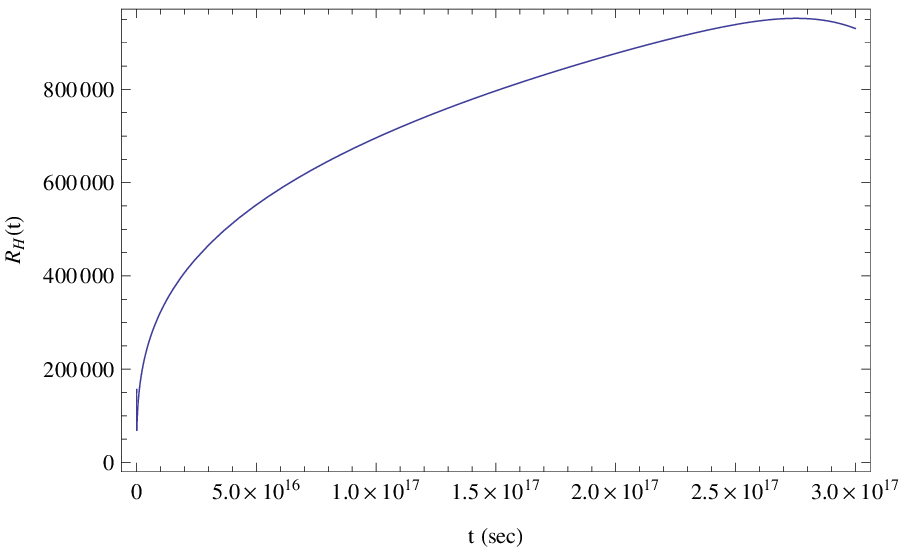}
\caption{The Hubble radius $R_H(t)$ as a function of the cosmic time, for $\gamma=2.1$ $\delta=2.5$, $t_0\simeq 10^{17}\mathrm{sec}$, $t_s\simeq 10^{-15}\mathrm{sec}$, $H_0\simeq 6.293\times 10^{13}\mathrm{sec}^{-1}$, $H_i\simeq 6\times 10^{26}\mathrm{sec}^{-1}$, $t_m=10^{12}$sec and for $f_0=10^{-97}\mathrm{sec}^{-\gamma-\delta-1}$ (left plot) and $f_0=10^{-98}\mathrm{sec}^{-\gamma-\delta-1}$ (right plot)}
\label{hubradiusnew1}
\end{figure}

Now we describe in brief the behavior of the Hubble horizon radius $R_H(t)=1/(aH)$, corresponding to the model (\ref{newmodel}), and we discuss the qualitative features that render the model a successful approximation of a viable cosmology. In Figs. \ref{hubradiusnew} and \ref{hubradiusnew1}, we have plotted the time dependence of the Hubble radius corresponding to the model (\ref{newmodel}), for three values of the parameter $f_0$, namely for $f_0=10^{-96}$ (Fig. \ref{hubradiusnew}), for $f_0=10^{-98}$ (left plot of Fig. \ref{hubradiusnew1}), and for $f_0=10^{-99}$ (Fig. \ref{hubradiusnew1}). 
Before analyzing the behavior, let us note that the resulting picture of the late-time behavior is crucially affected by the parameter $f_0$, as it can be seen from Fig. \ref{hubradiusnew1}. Actually as $f_0$ decreases, the deceleration-acceleration transition occurs at later and later times. So we focus on the case depicted in Fig. \ref{hubradiusnew}, so for $f_0=10^{-96}\mathrm{sec}^{-\gamma-\delta-1}$. For this case, the Hubble radius at early times decreases in a nearly exponential way (recall it is a quasi de Sitter accelerating era), then it increases during the radiation and matter domination era and it decreases again after the second Type IV singularity which occurs at $t\sim 10^{17}$sec. Clearly this is a viable cosmological behavior, when primordial quantum fluctuating modes are taken into account. Particularly, the primordial vacuum quantum modes are initially well inside the Hubble radius, at the beginning of the inflationary era, at which point their wavenumber satisfies $k\gg H(t)a(t)$. 
Then, during inflation the Hubble radius decreases and effectively the modes exit the Hubble radius when $k=aH$, and after that these modes go outside the horizon. After the inflationary era, and during the radiation and matter domination era, the Hubble radius increases again, and effectively the modes re-enter the horizon, and this is why these modes are relevant for present day observations. Then, the Hubble horizon increases until after the second Type IV singularity, where the deceleration to acceleration transition occurs, and the Hubble horizon starts to decrease again. Note that the behavior we just described characterizes a viable inflationary cosmology \cite{inflation1,inflation2,inflation3,inflation4,inflation5,inflation6,inflation7,inflation8}. In conclusion the model (\ref{newmodel}) has the following appealing qualitative features:
\begin{itemize}
    \item It describes in a unified way all the cosmological eras of the Universe, namely, the late and early-time acceleration era and the matter and radiation domination eras.
    \item It is predicted that there is an intermediate era between the radiation and matter domination transition, where the EoS of the Universe continuously deforms from a radiation dominated value to a matter domination value. Note that such an EoS describes collisional matter fluids \cite{collisional}. 
\item The Hubble horizon behaves in the correct way, as is expected for a viable cosmological evolution.
\end{itemize}    
What remains unanswered for the moment is the question if the power spectrum that is generated during inflation is scale invariant and secondly, if the primordial modes, after the horizon exit evolve or these are described by a constant stochastic mean value. These questions shall be addressed in detail in a later section of this article. Note that the answer to these questions strongly depends on the modified gravity model which realizes the cosmological evolution we just described, so these are of fundamental importance, since this will determine if the $F(G)$ gravity approach we shall use in this paper is consistent with observations. Before we answer these, in the next section we investigate how the cosmological evolution (\ref{newmodel}) can be realized in the context of $F(G)$ gravity.

\section{Unified Cosmological Evolution from $F(G)$ Gravity}

We now proceed to the realization of the unification cosmology (\ref{newmodel}) by vacuum $F(G)$ gravity. To this end we shall use well known reconstruction techniques \cite{reviews11,fg4,fg5,fg6,fg7}, and we investigate which $F(G)$ gravity can approximately generate each era of the model (\ref{newmodel}). Consider the following Jordan frame vacuum $F(G)$ gravity action, 
\begin{equation}\label{actionfggeneral}
\mathcal{S}=\frac{1}{2\kappa^2}\int \mathrm{d}^4x\sqrt{-g}\left ( R+F(G)\right )\, ,
\end{equation}
with $\kappa^2=1/M_{pl}^2$, and $M_{pl}=1.22\times 10^{19}$GeV. Upon variation of the action (\ref{actionfggeneral}) with respect to the metric, we obtain the following field equations,
\begin{align}\label{fgr1}
& R_{\mu \nu}-\frac{1}{2}g_{\mu \nu}F(G)-\Big{(}-2RR_{\mu \nu}+4R_{\mu \rho}R_{\nu}^{\rho}-2R_{\mu}^{\rho \sigma \tau}R_{\nu \rho \sigma \tau}+4g^{\alpha \rho}g^{\beta \sigma}R_{\mu \alpha \nu \beta}R_{\rho \sigma}\Big{)}F'(G)\\ \notag &
-2 \left (\nabla_{\mu}\nabla_{\nu}F'(G)\right )R+2g_{\mu \nu}\left (\square F'(G) \right )R-4 \left (\square F'(G) \right )R_{\mu \nu }+4 \left (\nabla_{\mu}\nabla_{\nu}F'(G)\right )R_{\nu}^{\rho }\\ \notag &+4 \left (\nabla_{\rho}\nabla_{\nu}F'(G)\right ) R_{\mu}^{\rho}
-4g_{\mu \nu} \left (\nabla_{\rho}\nabla_{\sigma }F'(G)\right )R^{\rho \sigma }+4 \left (\nabla_{\rho}\nabla_{\sigma }F'(G)\right )g^{\alpha \rho}g^{\beta \sigma }R_{\mu \alpha \nu \beta }=0
\end{align}
where $G$ is the Gauss-Bonnet invariant, which in terms of the Hubble rate reads, 
\begin{equation}\label{gausbonehub}
G=24H^2\left (\dot{H}+H^2 \right )\, .
\end{equation}
For a flat Friedmann-Robertson-Walker (FRW) metric of the form, 
\begin{equation}\label{metricformfrwhjkh}
\mathrm{d}s^2=-\mathrm{d}t^2+a^2(t)\sum_i\mathrm{d}x_i^2\, ,
\end{equation}
the field equations (\ref{fgr1}) become,
\begin{align}\label{eqnsfggrav}
& 6H^2+F(G)-GF'(G)+24H^3\dot{G}F''(G)=0\\ \notag &
4\dot{H}+6H^2+F(G)-GF'(G)+16H\dot{G}\left ( \dot{H}+H^2\right ) F''(G)
\\ \notag & +8H^2\ddot{G}F''(G)+8H^2\dot{G}^2F'''(G)=0\, .
\end{align}
By introducing the auxiliary functions $P(t)$ and $Q(t)$, the action (\ref{actionfggeneral}) becomes,
\begin{align}\label{actionfrg}
& \mathcal{S}=\frac{1}{2\kappa^2}\int \mathrm{d}^4x\sqrt{-g}\left ( R+P(t)G+Q(t)\right )\, ,
\end{align}
and upon variation with respect to $t$, we acquire the following equation,
\begin{equation}\label{auxeqnsvoithitiko}
\frac{\mathrm{d}P(t)}{\mathrm{d}t}G+\frac{\mathrm{d}Q(t)}{\mathrm{d}t}=0\, .
\end{equation}
In effect, by solving Eq. (\ref{auxeqnsvoithitiko}), with respect to $t=t(G)$, then by substituting the result in the following equation,
\begin{equation}\label{ebasc}
F(G)=P(t)G+Q(t)\, ,
\end{equation}
this will give the resulting $F(G)$ gravity which realizes the given Hubble rate $H(t)$. Practically, the vital element of the reconstruction technique is to find the functions $P(t)$ and $Q(t)$, so we now present how to find these in a direct way. Upon combining Eq. (\ref{ebasc}) with the first of the equations of Eq. (\ref{eqnsfggrav}), we get,
\begin{align}\label{ak}
& Q(t)=-6H^2(t)-24H^3(t)\frac{\mathrm{d}P}{\mathrm{d}t}\, ,
\end{align}
which yields the function $Q(t)$ as a function of $P(t)$. Then by substituting Eq. (\ref{ak}) in Eq. (\ref{ebasc}), we get the following differential equation,
\begin{align}\label{diffept}
& 2H^2(t)\frac{\mathrm{d}^2P}{\mathrm{d}t^2}+2H(t)\left (2\dot{H}(t)-H^2(t) \right )\frac{\mathrm{d}P}{\mathrm{d}t}+\dot{H}(t)=0\, ,
\end{align}
the solution of which gives the function $P(t)$. Finding $P(t)$, we also have $Q(t)$ and then we can obtain the function $t=t(G)$. Finally, upon substitution of the resulting expression in Eq. (\ref{ebasc}), we obtain the $F(G)$ gravity. In the following sections we make extensive use of this method for finding the $F(G)$ gravity that realizes each era.

\subsection{$F(G)$ Gravity at Early Times}

At early times, the Hubble rate is approximated by the expression appearing in Eq. (\ref{approxearlynew}), so the differential equation (\ref{diffept}) becomes approximately equal to,
\begin{align}\label{diffept2}
& \left(-\frac{2 H_0^2-4 H_0 H_i t_i}{H_i}\right)\frac{\mathrm{d}^2P}{\mathrm{d}t^2}+\left(4 H_0+\frac{2 H_0^3}{H_i}+6 H_0^2 t_i+4 H_i t_i\right)\frac{\mathrm{d}P}{\mathrm{d}t}+1=0\, ,
\end{align}
where we took into account that $H_0,H_i\gg t$, at early times. By solving the differential equation (\ref{diffept2}) we obtain the function $P(t)$,
\begin{equation}\label{dfesolu}
P(t)\simeq -\frac{H_i t}{2 \kappa_1}+\frac{C_1 \kappa_3 e^{\kappa_2 t}}{\kappa_1}+C_2\, ,
\end{equation}
where $C_1$ and $C_2$ are arbitrary integration constants and also the parameters $\kappa_1,\kappa_2,\kappa_3$ can be found in the Appendix. Accordingly, upon substituting Eq. (\ref{dfesolu}) in Eq. (\ref{ak}), we acquire the function $Q(t)$, which reads,
\begin{equation}\label{qt}
Q(t)\simeq \omega_1+\omega_2 t+\omega_3 e^{\kappa_2 t}\, ,
\end{equation} 
where the parameters $\omega_1,\omega_2,\omega_3$ can also be found in the Appendix. Then by substituting $P(t)$ and $Q(t)$ in Eq. (\ref{auxeqnsvoithitiko}), and by solving with respect to $t$ we get, 
\begin{equation}\label{finaformofauxiliaryeq}
t=\frac{\ln\left[\frac{G H_i-2 \kappa_1 \omega_2}{2 \kappa_2 (G \kappa_3+\kappa_1 \omega_3)}\right]}{\kappa_2}\, .
\end{equation}
So finally, the $F(G)$ gravity reads,
\begin{equation}\label{fgrgravity1}
F(G)\simeq \frac{G H_i+2 C_2 G \kappa_1 \kappa_2+2 \kappa_1 \kappa_2 \omega_1-2 \kappa_1 \omega_2+(-G H_i+2 \kappa_1 \omega_2) \ln\left[\frac{G H_i-2 \kappa_1 \omega_2}{2 G \kappa_2 \kappa_3+2 \kappa_1 \kappa_2 \omega_3}\right]}{2 \kappa_1 \kappa_2}\, .
\end{equation}
We can further simplify the resulting $F(G)$ gravity, by taking into account that the Gauss-Bonnet invariant (\ref{gausbonehub}) at early times is approximately equal to,
\begin{equation}\label{gearlytimes}
G\simeq 24 H_0^4-24 H_0^2 H_i-96 H_0^3 H_i t\, ,
\end{equation}
and since $H_0,Hi\gg 1$ and $t\ll 1$, this means that at early times, $G\rightarrow \infty$. So the resulting $F(G)$ gravity at leading order is,
\begin{equation}\label{fgearly}
F(G)\simeq \rho_1 G-\frac{\rho_3}{G} +\rho_2\, ,
\end{equation}
where the parameters $\rho_1,\rho_2,\rho_3$ can be found in the Appendix. The approximate expression for the $F(G)$ gravity shall play a crucial role in the calculation of the power spectrum of the primordial curvature perturbations, so we will make extensive use of Eq. (\ref{fgearly}) in the sections to follow.

In a similar way, we can find the approximate expressions for the $F(G)$ gravities that realize the rest cosmological eras, and we omit most of the details for brevity. Particularly, for the radiation domination era, in which case the Hubble rate is approximately equal to $H(t)\simeq \frac{1}{2t}$, so the resulting $P(t)$ function is,
\begin{equation}\label{ptfunction}
P(t)=-\frac{t^2}{3}+\frac{2}{7} t^{7/2} C_4+C_3\, ,
\end{equation}
where $C_3,C_4$ are arbitrary integration constants. By choosing for simplicity $C_4=0$, the function $Q(t)$ reads,
\begin{equation}\label{qtf}
Q(t)=\frac{7}{4 t^2}\, ,
\end{equation}
and therefore the function $t(G)$ reads,
\begin{equation}\label{tgfunction}
t(G)=\frac{(21)^{1/4}}{\sqrt{2} G^{1/4}}\, .
\end{equation}
Consequently, the resulting $F(G)$ gravity is,
\begin{equation}\label{fg}
F(G)\simeq C_3\, G\, .
\end{equation}
Before proceeding to the matter domination era, we need to note that in the case of $F(R)$ gravity description of the radiation domination era we needed to find leading order approximations of the Hubble rate, since in the case that the Hubble rate is exactly given by $H(t)\simeq \frac{1}{2t}$, the Ricci scalar was equal to zero, see \cite{nooa} for more details on this leading order expansion. In the Gauss-Bonnet description however, this problem no longer persists, since in this case the Gauss-Bonnet invariant is $G=-\frac{3}{2 t^4}$, which is non-zero.

We now proceed to the matter domination era $F(G)$ gravity description, with the Hubble rate being in this case equal to $H(t)\simeq \frac{2}{3t}$, so the resulting $P(t)$ function is,
\begin{equation}\label{ptfunction}
P(t)=-\frac{9 t^2}{40}+\frac{3}{11} t^{11/3} C_5+C_6\, ,
\end{equation}
where $C_5,C_6$ are again arbitrary integration constants. By choosing for simplicity $C_5=0$, the function $Q(t)$ reads in this case,
\begin{equation}\label{qtf}
Q(t)=\frac{124}{45 t^2}\, ,
\end{equation}
and consequently the function $t(G)$ reads,
\begin{equation}\label{tgfunction}
t(G)=\frac{2 (62)^{1/4}}{3 G^{1/4}}\, ,
\end{equation}
and as in the radiation domination case, the resulting $F(G)$ gravity description reads,
\begin{equation}\label{fg}
F(G)\simeq C_6\, G\, ,
\end{equation}
which has the same power dependence with respect to the Gauss-Bonnet invariant as in the radiation domination case. Finally, following the same procedure, the resulting $F(G)$ gravity that realizes the late-time behavior is,
\begin{equation}\label{latetimefg}
F(G)\simeq \mu_1 e^{\mu_2\text{  }G^{\xi }} G^{\varepsilon }\, ,
\end{equation}
where the parameters $\mu_1,\mu_2,\xi,\varepsilon$, can be found in the Appendix. In Table \ref{table1}, we gathered the resulting expressions for the $F(G)$ gravity in all the cosmological eras.
\begin{table*}[h]
    \small
\begin{tabular}{@{}|c|r|rrrrrrrrrr@{}}
        \tableline
        \tableline
        \tableline
        Cosmological Era & $F(G)$ Gravity Realization 
        \\\tableline
        Early-time era &  $F(G)\simeq \rho_1 G-\frac{\rho_3}{G} +\rho_2$
        \\\tableline
        Radiation Domination Era & $F(G)\simeq C_3\, G$ 
        \\\tableline
        Matter Domination Era & $F(G)\simeq C_6\, G$
        \\\tableline
        Late-time era & $F(G)\simeq \mu_1 e^{\mu_2\text{  }G^{\xi }} G^{\varepsilon }$ 
        \\\tableline
        \tableline
    \end{tabular}
    \caption{\label{table1}The $F(G)$ gravities which realize the cosmological eras of the Universe. The various parameters can be found in the Appendix.}
\end{table*}

\section{Scalar Perturbations and the Power Spectrum of Primordial Curvature Perturbations}

As we demonstrated in the previous sections, the $F(G)$ gravity theoretical framework can successfully produce the cosmological eras corresponding to the evolution (\ref{newmodel}) and now the question is if the corresponding evolution is compatible with the observational data. As was shown in \cite{nooa}, in the context of $F(R)$ gravity the cosmological evolution was compatible with the observational data, since the early-time $F(R)$ gravity was described by the $R^2$ Starobinsky inflation model, so the primordial cosmological perturbations have a nearly scale invariant power spectrum. In this section we explicitly calculate in detail the power spectrum of primordial curvature perturbations, and as we demonstrate, the resulting power spectrum is not scale invariant, indicating that not all modified gravity realizations of the same cosmological evolution are successful when confronted with observational data. In addition, another interesting issue which we address in this section has to do with the evolution of the primordial curvature perturbations that exit the Hubble horizon at early times. Particularly the question is if these evolve in time after the initial horizon crossing. In the $F(R)$ gravity description, once these modes exited the horizon during inflation, they froze and therefore were described by a stochastic mean value with no time dependence. However, this is not granted for any cosmological evolution, for example in matter bounce cosmologies the perturbations that exited the horizon evolve in time, see \cite{branderberger,branderberger1,branderberger2}. As we demonstrate in the $F(G)$ gravity case, these modes decay exponentially in time.

\subsection{Explicit Calculation of the Power Spectrum of Primordial Curvature Perturbations}

We shall calculate the power spectrum of scalar primordial curvature perturbations, so we consider the following linear scalar perturbations of the FRW background (\ref{metricformfrwhjkh}), 
\begin{equation}\label{scalarpertrbubations}
\mathrm{d}s^2=-(1+\psi)\mathrm{d}t^2-2a(t)\partial_i\beta\mathrm{d}t\mathrm{d}x^i+a(t)^2\left( \delta_{ij}+2\phi\delta_{ij}+2\partial_i\partial_j\gamma \right)\mathrm{d}x^i\mathrm{d}x^j\, ,
\end{equation}
where the functions $\psi$, $\phi$, $\gamma$ and $\beta$ quantify the smooth scalar perturbations of the FRW metric. By using the linearly perturb FRW metric (\ref{scalarpertrbubations}), the calculation of the power spectrum is straightforward and for details we refer to \cite{inflation1,inflation2,inflation3,inflation4,inflation5,inflation6,inflation7,inflation8,noh,branderberger,felice}. The perturbations can be evaluated quantitatively in an elegant way if gauge invariant quantities are considered, so we shall make use of the following gauge invariant quantity,
\begin{equation}\label{confedf}
\Phi=\phi-\frac{H\delta \xi}{\dot{\xi}}\, ,
\end{equation}
which is known as the comoving curvature perturbation, where the function $\xi$ stands for $\xi=\frac{\mathrm{d}F}{\mathrm{d}G}$. The master equation that determines the evolution of the scalar perturbation modes in the context of $F(G)$ gravity is \cite{noh,felice},
\begin{equation}\label{perteqnmain}
a(t)^3\mathcal{Q}(t)\ddot{\Phi}+\left(3a(t)^2\dot{a}\mathcal{Q}(t)+a(t)^3\dot{\mathcal{Q}}(t)\right)\dot{\Phi}+B_1(t)\mathcal{Q}(t)a(t)k^2\Phi=0\, ,
\end{equation}
and it can be seen that the $k^2$ modes determine the evolution at early times. The function $B_1(t)$ corresponds to the propagation speed and in the $F(G)$ gravity case it is equal to,
\begin{equation}\label{b1}
B_1(t)=1+\frac{2\dot{H}}{H^2}\, ,
\end{equation}
while the function $\mathcal{Q}(t)$ in Eq. (\ref{perteqnmain}) is equal to,
\begin{equation}\label{gfgdhbhhyhjs}
\mathcal{Q}(t)=\frac{6\left( \frac{\mathrm{d}^2F}{\mathrm{d}G^2}\right )^2\dot{G}^2\left(1+4F''(G)\dot{G}H\right)}{\left(1+6HF''(G)\dot{G}\right)^2}\, ,
\end{equation}
where $\dot{\xi}=\frac{\mathrm{d}F^2}{\mathrm{d}G^2}\dot{G}$, and with the dot denoting as usual differentiation with respect to the cosmic time, while the prime differentiation with respect to the Gauss-Bonnet scalar $G$. We are interested in finding an approximation of the power spectrum at early times, so the $F(G)$ gravity is approximately equal to the expression given in Eq. (\ref{fgearly}), while the Hubble rate at early times is given in Eq. (\ref{approxearlynew}). By taking into account that $H_0,H_i\gg t$ at early times, the differential equation that determines the evolution of primordial curvature perturbations (\ref{perteqnmain}) becomes approximately equal to,
\begin{align}\label{approximatediff}
& \Big{(}\frac{H_i^2 \left(-2 H_0^2+H_i\right)^2 \rho_3^2 \left(1+\frac{\left(2 H_0^2-H_i\right) H_i \rho_3}{36 H_0 \left(H_0^3-H_0 H_i\right)^3}\right) }{3456 H_0^4 \left(H_0^3-H_0 H_i\right)^6 \left(1+\frac{\left(2 H_0^2-H_i\right) H_i \rho_3}{24 H_0 \left(H_0^3-H_0 H_i\right)^3}\right)}\Big{)}\ddot{\Phi}+\Big{(}\frac{H_i^2 \left(-2 H_0^2+H_i\right)^2 (H_0-H_i (t-t_i)) \rho_3^2 \left(1+\frac{\left(2 H_0^2-H_i\right) H_i \rho_3}{36 H_0 \left(H_0^3-H_0 H_i\right)^3}\right) }{1152 H_0^4 \left(H_0^3-H_0 H_i\right)^6 \left(1+\frac{\left(2 H_0^2-H_i\right) H_i \rho_3}{24 H_0 \left(H_0^3-H_0 H_i\right)^3}\right)}\Big{)}\dot{\Phi}\\ \notag &
\Big{(}\frac{H_i^2 \left(-2 H_0^2+H_i\right)^2 \left(1-\frac{2 H_i}{H_0^2}\right) k^2 \rho_3^2 \left(1+\frac{\left(2 H_0^2-H_i\right) H_i \rho_3}{36 H_0 \left(H_0^3-H_0 H_i\right)^3}\right) }{3456 H_0^4 \left(H_0^3-H_0 H_i\right)^6 \left(1+\frac{\left(2 H_0^2-H_i\right) H_i \rho_3}{24 H_0 \left(H_0^3-H_0 H_i\right)^3}\right)}\Big{)}\Phi=0\, ,
\end{align}
and by solving it, we obtain the following expression which describes the evolution of the comoving perturbations at early times,
\begin{equation}\label{solutionevolutionnewpap}
\Phi(t)= e^{\frac{\left(-3 H_0^3-\sqrt{9 H_0^6-4 H_0^4 k^2+8 H_0^2 H_i k^2}\right) t}{2 H_0^2}} \mathcal{C}_1+e^{\frac{\left(-3 H_0^3+\sqrt{9 H_0^6-4 H_0^4 k^2+8 H_0^2 H_i k^2}\right) t}{2 H_0^2}} \mathcal{C}_2\, ,
\end{equation}
where $\mathcal{C}_1,\mathcal{C}_2$ are arbitrary integration constants. Note that since we are interested in finding the behavior at early times, this corresponds to the large wavenumber limit, that is $k\gg a H$, so in this limit, the comoving perturbation (\ref{solutionevolutionnewpap}) becomes,
\begin{equation}\label{solutionevolution}
\Phi(t)= \left(\frac{\mathcal{C}_2 \sqrt{-4 H_0^4+8 H_0^2 H_i}}{2 H_0^2}+\frac{\mathcal{C}_2 \sqrt{-4 H_0^4+8 H_0^2 H_i}}{2 H_0^2}\right) k t\, ,
\end{equation}
where for simplicity we chose the integration constants to satisfy $\mathcal{C}_1=-\mathcal{C}_2$. Having the time dependence of the comoving curvature perturbation at hand, we can easily compute the power spectrum, which in terms of $\Phi$ is equal to,
\begin{equation}\label{powerspecetrumfgr}
\mathcal{P}_R=\frac{4\pi k^3}{(2\pi)^3}|\Phi|_{k=aH}^2\, ,
\end{equation}
and it is defined at the horizon crossing. So by substituting Eq. (\ref{solutionevolution}) to (\ref{powerspecetrumfgr}), the power spectrum reads,
\begin{equation}\label{powerspectrajb}
\mathcal{P}_R\sim k^3 \Big{|} \left(\frac{\mathcal{C}_2 \sqrt{-4 H_0^4+8 H_0^2 H_i}}{2 H_0^2}+\frac{\mathcal{C}_2 \sqrt{-4 H_0^4+8 H_0^2 H_i}}{2 H_0^2}\right) k t\Big{|}^2_{k=aH}\, ,
\end{equation} 
and this has to be evaluated at the horizon crossing. By taking into account the horizon crossing condition $k=aH$, the power spectrum becomes,
\begin{equation}\label{hcpower1}
\mathcal{P}_R\sim \frac{\left(\frac{\mathcal{C}_2 \sqrt{-4 H_0^4+8 H_0^2 H_i}}{2 H_0^2}+\frac{\mathcal{C}_2 \sqrt{-4 H_0^4+8 H_0^2 H_i}}{2 H_0^2}\right) k(H_0-k)}{2 H_0^2 H_i}\, ,
\end{equation}
so what remains is to calculate the $k$-dependence of the integration constant $\mathcal{C}_2$, which can be easily done by taking into account the initial conditions for the comoving curvature perturbation. Particularly we assume that the comoving perturbation $\Phi$ is related to a Bunch-Davies initial vacuum state, in a way we now demonstrate. For convenience we use a canonical scalar field $u$ and the conformal time $\tau$, which is related to the cosmic time as $\mathrm{d}\tau =a^{-1}(t)\mathrm{d}t$. At early times, since $t\ll 1$, the exponential scale factor describing the nearly de Sitter evolution is approximately equal to $a\simeq 1$, so practically $\tau$ and $t$ at early times are approximately the same. The canonical scalar $u$ is defined in terms of the comoving curvature $\Phi$ as $u=z_s \Phi$, where $z_s=Q(t)a(t)$, so at early times we have,
\begin{equation}\label{safakebelieve}
u\sim \Phi \mathcal{Q}(t)\, ,
\end{equation}
where $\mathcal{Q}(t)$ is given in Eq. (\ref{gfgdhbhhyhjs}). The action corresponding to the canonical scalar field $u$ is equal to,
\begin{equation}\label{actiaonerenearthebounce}
\mathcal{S}_u\simeq \int \mathrm{d}^3\mathrm{d}\tau \left[ \frac{u'}{2}-\frac{1}{2}(\nabla u)^2+\frac{z_s''}{z_s}u^2\right ]\, ,
\end{equation}
with the prime this time indicating differentiation with respect to the conformal time, which at early times is approximately identical to the cosmic time. We therefore assume that the canonical scalar is identical to a Bunch-Davies vacuum state before inflation, so $u\sim \frac{e^{-ik\tau}}{\sqrt{k}}$, at early times, and hence the integration constant $\mathcal{C}_2$ is approximately equal to,
\begin{equation}\label{constantc2}
\mathcal{C}_2=\frac{1}{\sqrt{k}\frac{H_i^2 \left(-2 H_0^2+H_i\right)^2 \rho_3^2 \left(1+\frac{\left(2 H_0^2-H_i\right) H_i \left(H_0+H_i \left(-\frac{H_0-k}{H_i}+t_i\right)\right) \rho_3}{36 H_0^2 \left(H_0^3-H_0 H_i\right)^3}\right)}{3456 H_0^4 \left(H_0^3-H_0 H_i\right)^6 \left(1+\frac{\left(2 H_0^2-H_i\right) H_i \rho_3}{24 H_0 \left(H_0^3-H_0 H_i\right)^3}\right)}\left(\frac{ \sqrt{-4 H_0^4+8 H_0^2 H_i}}{2 H_0^2}+\frac{ \sqrt{-4 H_0^4+8 H_0^2 H_i}}{2 H_0^2}\right) t_i}\, ,
\end{equation}
so by combining the above results we obtain the $k$-dependence of the power spectrum $\mathcal{P}_R$  at early times,
\begin{equation}\label{powerspectrumfinal}
\mathcal{P}_R\sim k^3 \left(\frac{1728 H_0^2 \left(H_0^3-H_0 H_i\right)^6 (H_0-k) \sqrt{k} \left(1+\frac{\left(2 H_0^2-H_i\right) H_i \rho_3}{24 H_0 \left(H_0^3-H_0 H_i\right)^3}\right)}{H_i^3 \left(-2 H_0^2+H_i\right)^2 t_i \rho_3^2 }\right)^2\, ,
\end{equation}
which is not scale invariant, but the question is if it can be nearly scale invariant. In order to see this, we now calculate the spectral index of the primordial curvature perturbations, which is defined in terms of the power spectrum $\mathcal{P}_R$ as follows, 
\begin{equation}
 n_s-1\equiv\frac{d\ln\mathcal{P}_{\mathcal{R}}}{d\ln
k}\, 
\end{equation}
so by inserting the result of Eq. (\ref{powerspectrumfinal}), the resulting spectral index reads,
\begin{align}\label{spectralindexresult}
& n_s \simeq -\frac{\frac{243}{512} H_i^{11/2} \left(\frac{1}{2} \sqrt{H_i} \left(4+k^2\right)-k \left(6+k^2\right)\right)}{k^2 \left(-\frac{\sqrt{H_i}}{2}+k\right) \left(\frac{243 H_i^{11/2}}{512}+\frac{1}{2} H_i^2 (k+H_i t_i) \rho_3\right)}\\ \notag &
-\frac{\frac{1}{2} H_i^2 \left(\frac{1}{2} \sqrt{H_i} \left(k \left(2+k^2\right)+H_i \left(4+k^2\right) t_i\right)-k \left(k \left(4+k^2\right)+H_i \left(6+k^2\right) t_i\right)\right) \rho_3}{k^2 \left(-\frac{\sqrt{H_i}}{2}+k\right) \left(\frac{243 H_i^{11/2}}{512}+\frac{1}{2} H_i^2 (k+H_i t_i) \rho_3\right)}
\, ,
\end{align}
and it can be checked that the resulting spectrum cannot produce a spectral index compatible the recent Planck observational data \cite{planck1,planck2}. Therefore, the comoving curvature perturbations corresponding to the $F(G)$ gravity theory do not produce a scale invariant spectrum. Before closing, note that the resulting picture might be an artifact of the approximate expressions we used for obtaining the early-time $F(G)$ gravity. Hence, it could be possible that a Loop Quantum Cosmology corrected $F(G)$ gravity theory \cite{sergeiharooikonomou} could produce a scale invariant spectrum. This can produce however a red spectrum, so it might be doubtful how the quantum corrections could alter the resulting picture. We hope to address this issue in detail in the future.

\subsection{Evolution of the Primordial Curvature Perturbations after the Horizon Crossing}

A last issue we want to address before we close thus section, is the evolution of the primordial perturbations after these exit the Hubble horizon during inflation. In standard inflationary cosmology, the primordial perturbations do not evolve in time after the horizon crossing and these are described by a stochastic mean value, and hence these are conserved and remain unaltered until these enter the horizon at a later stage. This is a crucial feature of the standard inflationary scenario, since the conservation of the primordial modes enables us to make predictions about inflation. This is due to the fact that when these modes reenter the horizon, these reveal us the form of perturbations when these exited the horizon at early times. There are scenarios in cosmology for which the perturbations evolve in time, like for example in the matter bounce scenario case \cite{branderberger}. As we now demonstrate, the primordial perturbations do not grow in the present case, but exponentially decay at times after the horizon crossing.

In order to proceed, we will focus on primordial modes for which the wavenumber satisfies $k\ll a(t)H(t)$. In that case, the differential equation (\ref{perteqnmain}) becomes,
\begin{equation}\label{perteqnmainportoriko}
\frac{1}{a(t)^3\mathcal{Q}(t)}\frac{\mathrm{d}}{\mathrm{d}t}\left(a(t)^3\mathcal{Q}(t)\dot{\Phi}\right)=0\, ,
\end{equation} 
the solution of which is,
\begin{equation}\label{soldiffeqn}
\Phi (t)=\mathcal{C}_1+\mathcal{C}_2\int \frac{1}{a(t)^3\mathcal{Q}(t)}\mathrm{d}t\, ,
\end{equation}
where $\mathcal{Q}(t)$ is defined in Eq. (\ref{gfgdhbhhyhjs}), and $\mathcal{C}_1,\mathcal{C}_2$ are arbitrary integration constants. We shall be interested in cosmic times just after the end of inflation, so the scale factor can be approximated by $a\sim e^{H_0 t}$, and also $\mathcal{Q}(t)$ is approximately equal to,
\begin{equation}\label{mathcalq}
\mathcal{Q}\simeq \frac{H_i^2 \left(-2 H_0^2+H_i\right)^2 \rho_3^2 \left(1+\frac{\left(2 H_0^2-H_i\right) H_i (H_0) \rho_3}{36 H_0^2 \left(H_0^3-H_0 H_i\right)^3}\right)}{3456 H_0^4 \left(H_0^3-H_0 H_i\right)^6 \left(1+\frac{\left(2 H_0^2-H_i\right) H_i (H_0) \rho_3}{24 H_0^2 \left(H_0^3-H_0 H_i\right)^3}\right)}\, ,
\end{equation}  
where we kept only dominant terms. So by inserting $\mathcal{Q}$ and $a(t)$ in Eq. (\ref{soldiffeqn}), we acquire,
\begin{equation}\label{finalresultforpertevol}
\Phi (t)\simeq \mathcal{C}_1+\mathcal{C}_2 \mathcal{A} e^{-H_0 t}\, ,
\end{equation}
where the parameter $\mathcal{A}$ is,
\begin{equation}\label{mathcala}
\mathcal{A}=\frac{3456\text{  }H_0^3 \left(H_0^3-H_0 H_i\right)^6 \left(1+\frac{\left(2 H_0^2-H_i\right) H_i \rho_3}{24 H_0^4 \left(H_0^2-H_i\right)^3}\right)}{H_i^2 \left(2 H_0^2-H_i\right)^2 \rho_3^2 \left(1+\frac{\left(2 H_0^2-H_i\right) H_i \rho_3}{36 H_0^4 \left(H_0^2-H_i\right)^3}\right)}\, .
\end{equation}
Therefore, the functional form of the evolution (\ref{finalresultforpertevol}) indicates that the evolution of the perturbations much later than the horizon crossing exponentially decay. We need to note that the behavior we just described is very common in all the inflationary scenarios, so this result was in some sense expected, so the $F(G)$ case generates primordial perturbations which are conserved after the horizon crossing.

\section{Concluding Remarks}

In this article we investigated which vacuum $F(G)$ gravity can realize a cosmological evolution which can approximately describe the four evolution eras of our Universe. Particularly, we were interested in finding which $F(G)$ gravity can realize each cosmological era, and we calculated the power spectrum of primordial curvature perturbations corresponding to early-time evolution. As we demonstrated, the $F(G)$ gravity description differs from the $F(R)$ gravity description, because the resulting power spectrum is not compatible with the latest observational data. Thus, this quantitatively proves that not all modified gravity descriptions are equivalent and successful, when these are confronted with observations. Of course, this is not obvious without explicitly calculating the power spectrum corresponding to each modified gravity description, so it would be interesting to address the problem we studied in this paper, in the context of more general modified gravity theories, like for example $F(R,G)$ gravity \cite{fg5,fg6,felice,frg,cappofrg}. In that case, as was demonstrated in \cite{felice}, $k^4$-powers of the wavenumber appear in the master equation which governs the power spectrum of primordial curvature perturbations, and thus superluminal modes appear in the spectrum. This behavior could alter the resulting picture for the cosmological evolution we studied in this paper, and we intend to address in  a future work. In the same fashion, the $F(T)$ gravity \cite{caisaridakis,nashed} case should also be investigated.

An interesting question is whether the holonomy corrected $F(G)$ theory \cite{LQC1,LQC2,LQC3,LQC4,LQC5} can make the power spectrum nearly scale invariant. Particularly, if we include the Loop Quantum Cosmology effects, the question is whether the power spectrum is rendered scale invariant. This problem could be addressed in the way it was treated in \cite{sergeiharooikonomou}, so the calculation of the power spectrum for the holonomy corrected theory is straightforward.

\section*{Acknowledgments}

This work is supported by Min. of Education and Science of Russia (V.K.O).

\section*{Appendix A: Detailed Form of the Parameters Used in the Text}

Here we present the detailed form of the various parameters we used in the text. The parameters $\kappa_1,\kappa_2,\kappa_3$ appearing in Eq. (\ref{dfesolu}) are equal to,
\begin{align}\label{kappaparameters}
& \kappa_1=\frac{1}{\left(H_0^3+2 H_0 H_i+3 H_0^2 H_i t_i+2 H_i^2 t_i\right)} \\ \notag &
\kappa_2=\frac{\left(H_0^3+2 H_0 H_i+3 H_0^2 H_i t_i+2 H_i^2 t_i\right)}{H_0 (H_0+2 H_i t_i)}\\ \notag &
\kappa_3=H_0 (H_0-2 H_i t_i)\, .
\end{align}
Also the parameters $\omega_1,\omega_2,\omega_3$ appearing in Eq. (\ref{qt}), are equal to,
\begin{align}\label{omegapars}
& \omega_1=-H_0^2+2 H_0 H_i t_i+\frac{12 H_0^3 H_i}{\kappa_1}+\frac{36 H_0^2 H_i^2 t_i}{\kappa_1}+\frac{36 H_0 H_i^3 t_i^2}{\kappa_1}\\ \notag &
\omega_2=2 H_0 H_i+2 H_i^2t_i-\frac{36 H_0^2 H_i^2}{\kappa_1}-\frac{72 H_0 H_i^3 t_i}{\kappa_1}\\ \notag &
\omega_3=\frac{24 H_0^3 \kappa_2 \kappa_3}{\kappa_1}+\frac{72 H_0^2 H_i t_i \kappa_2 \kappa_3}{\kappa_1}+\frac{72 H_0 H_i^2 t_i^2 \kappa_2 \kappa_3}{\kappa_1}-\frac{72\text{  }H_i^3 t t_i^2 \kappa_2 \kappa_3}{\kappa_1}+\frac{24 H_i^3 t_i^3 \kappa_2 \kappa_3}{\kappa_1}\, ,
\end{align}
and also the parameters $\rho_1,\rho_2,\rho_3$ appearing in Eq. (\ref{fgearly}) are equal to,
\begin{align}\label{rhopars}
& \rho_1=\frac{\left(H_i+2 C_2 \kappa_1 \kappa_2-H_i \ln\left[\frac{H_i}{2 \kappa_2 \kappa_3}\right]\right)}{2 \kappa_1 \kappa_2}
\\ \notag &
\rho_2=\left(\omega_1+\frac{H_i \omega_3}{2 \kappa_2 \kappa_3}+\frac{\omega_2 \ln\left[\frac{H_i}{2 \kappa_2 \kappa_3}\right]}{\kappa_2}\right) 
\\ \notag &
\rho_3=\frac{\kappa_1 (2 \kappa_3 \omega_2+H_i \omega_3)^2}{4 \left(H_i \kappa_2 \kappa_3^2\right)}\, .
\end{align}
Finally, the parameters $\mu_1,\mu_2,\xi,\varepsilon$, appearing in Eq. (\ref{latetimefg}) are equal to,
\begin{align}\label{rhopars}
& \mu_1=-2^{3-\frac{2}{1+6 (2 \gamma +2 \delta )}} 3^{1-\frac{1}{1+6 (2 \gamma +2 \delta )}} f_0^{\frac{3 (1+4 (2 \gamma +2 \delta ))}{1+6 (2 \gamma +2 \delta )}}\, ,
\\ \notag &
\mu_2=2^{-\frac{5+6 (2 \gamma +2 \delta )}{1+6 (2 \gamma +2 \delta )}} 3^{-\frac{2}{1+6 (2 \gamma +2 \delta )}} f_0^{-\frac{7+6 (2 \gamma +2 \delta )}{1+6 (2 \gamma +2 \delta )}}\, ,
\\ \notag &
\xi_3=\frac{2}{1+6 (2 \gamma +2 \delta )}\, ,
\\ \notag &
\varepsilon =\frac{6 (2 \gamma +2 \delta )}{1+6 (2 \gamma +2 \delta )}\, .
\end{align}

\end{document}